\documentstyle[aps,preprint]{revtex}
\def\.{\!\cdot\!}
\def\:{\cdots}
\def\[{\left[}
\def\]{\right]}
\def\({\left(}
\def\){\right)}

\def\bi{\begin{itemize}}

\def\ei{\end{itemize}}
\def\be{\begin{eqnarray}}
\def\ee{\end{eqnarray}}
\def\bn{\begin{enumerate}}
\def\en{\end{enumerate}}
\def\h{{1\over 2}}

\def\nn{\nonumber}

\def\r2{\sqrt{2}}

\def\x{\times}
\def\labels#1{\label{#1}}
\def\ket#1{|#1\rangle}
\def\eq#1{(\ref{#1})}

\def\rr2{{1\over\sqrt{2}}}

\begin{document}
\title{A 2-3 Symmetry in Neutrino Oscillations}
\author{C.S. Lam}
\address{Department of Physics, McGill University\\
3600 University St., Montreal, QC, Canada H3A 2T8\\
Email: Lam@physics.mcgill.ca}
\maketitle

\begin{abstract}
Maximum mixing in atmospheric neutrino oscillation, as well as vanishing of
the MNS matrix element  $U_{e3}$, are consequences of a 2-3 symmetry,
under which the neutrino mass matrix is invariant under the interchange
of second and third generation neutrinos. These predictions of the
2-3 symmetry are consistent with the results of 
Super-Kamiokande, K2K, and CHOOZ experiments. If the symmetry is
exact at a high-energy scale set by right-handed neutrinos, 
a deviation from these predictions generated by renormalization-group
corrections will occur at experimental energies. With an 
 MSSM dynamics, the result
can be made to agree with a global fit of the neutrino
data, if normal hierarchy is assumed on the neutrino mass spectrum and if
the mass of the electron-neutrino is at least about 0.025 eV. 
The presence of this mass lower bound is a novel and 
interesting feature of the symmetry that can be falsified by future
experiments. Of the three viable solar neutrino solutions,
only LMA gives a sizable MNS matrix element $U_{e3}$ that can hopefully
be detected in future reactor experiments.
Inverted neutrino mass hierarchy is not permitted by this symmetry.
\end{abstract}
%\pacs{11.80.Fv,11.15.Bt,11.15.-q}
\narrowtext

%\Large

%\section{Introduction}
In the basis where the mass matrix of the charged leptons is diagonal,
the left-handed flavor-based (symmetric)
neutrino mass matrix $m'$ is related to its diagonal form
$m=diag[m_1,m_2,m_3]$ by 
\be
u^Tm'u=m,\labels{mpum}\ee
where $u$ is the unitary MNS mixing matrix \cite{MNS},
conventionally parametrized as
\be
u=\pmatrix{c_1c_3&s_1c_3&s_3\cr
-s_1c_2-c_1s_2s_3&c_1c_2-s_1s_2s_3&s_2c_3\cr
s_1s_2-c_1c_2s_3&-c_1s_2-s_1c_2s_3&c_2c_3\cr}.\labels{u}\ee
The phases in $m_i$ and $u$ are not measurable in present experiments so
we shall assume them to be zero. The parameters in \eq{u} are related
to the mixing angles $\theta_1\equiv \theta_{12}, \theta_2\equiv\theta_{23}$,
and $\theta_3\equiv\theta_{13}$ by
$s_i=\sin\theta_i$ and $c_i=\cos\theta_i$. These angles
are controlled respectively by the
solar, atomospheric, and reactor neutrino oscillations.
As in Ref.~\cite{GGMPV}, we take all angles to be between 0 and 
$\pi/2$. In this article we assume sterile neutrinos to play no role in
these oscillations.

Over the years, there have been many attempts to understand the
texture of $m'$.
In this paper we assume $m'$ to have an exact symmetry under the
interchange of the second and third generation neutrinos, and study
its consequences. We shall refer to this symmetry from now on as the
2-3 symmetry. It is a symmetry in which
the matrix elements of $m'$ is invariant under the 
interchange of the flavor basis vectors $\ket{e}\leftrightarrow\ket{e}$ and 
$\ket{\mu}\leftrightarrow -\ket{\tau}$. The minus sign here 
is needed to keep the convention of having only positive mixing angles.
With this symmetry,
\be
m'_{e\mu}&=&-m'_{e\tau},\nn\\
m'_{\mu\mu}&=&m'_{\tau\tau},\labels{23sym}\ee
so we may parametrize $m'$ by four parameters as in
\be
m'=\pmatrix{a&b&-b\cr b&f&e\cr -b&e&f\cr}.\labels{mp}\ee
The second constraint in \eq{23sym} has previously been studied by
several authors \cite{EL,A23}.

The mass spectra of quarks and charged leptons suggest that if 
a generation symmetry is to hold approximately, then
it is much more natural to have a 1-2 symmetry \cite{A12} rather than
a 2-3 symmetry, because the third-generation fermion is much
heavier than those in the first two. For neutrinos, only mass differences
are known, but since $(\Delta m_{12})^2\ll (\Delta m_{23})^2$,
a 1-2 symmetry is still more natural
if the normal hierarchy $m_1<m_2<m_3$ holds with
$m_1\le|\Delta m_{12}|$. For this reason,
the 2-3 symmetry in \eq{23sym} might seem to be totally
unnatural. Nevertheless, it turns out that
the 2-3 symmetry on $m'$ given by \eq{23sym} places
 no restriction whatsoever
on the neutrino masses $m_i$, so this  2-3 symmetry on $m'$ 
is not contradictory to a 1-2 symmetry
on the mass spectrum. Moreover, the 2-3 symmetry gives rise naturally to 
the right mixing of neutrinos, as we shall see immediately.

The two conditions in \eq{23sym} implies
$s_3=0$ and $s_2=1/\sqrt{2}$, and an
MNS matrix \eq{u} equal to
\be
u&=&\pmatrix{c_1&s_1&0\cr -\rr2 s_1& \rr2 c_1&\rr2\cr
\rr2 s_1&-\rr2 c_1&\rr2\cr}=v\.w\nn\\
&\equiv&\pmatrix{1&0&0\cr 0&\rr2&\rr2\cr 0&-\rr2&\rr2\cr}
\.\pmatrix{c_1&s_1&0\cr -s_1&c_1&0\cr 0&0&1\cr}.\labels{u0}\ee
To derive this result, note that
\be
m''\equiv v^Tm'v=\pmatrix{a&\sqrt{2}b&0\cr \sqrt{2}b&f-e&0\cr
0&0&f+e\cr},\ee
and that $m''$ can be diagonalized by $w$. Incidentally,
the eigenvalues are $f+e$ and $\h[a+f-e\pm\sqrt{(a-f+e)^2+8b^2}]$,
and the solar mixing angle is determined by the positive solutions of
$\cot\theta_1=[f-e-a\pm\sqrt{(f-e-a)^2+8b^2}]/2\sqrt{2}b$, but these
relations are not needed in the following.

The MNS mixing matrix \eq{u0} impies maximal mixing in atmospheric
neutrino oscillations, consistent with the Super-Kamiokande
\cite{SK}  and the 
K2K \cite{K2K} observations. It also gives rise to
$u_{e3}=0$, consistent with the CHOOZ reactor experiment \cite{CHOOZ}. 
On the other hand, the solar mixing angle $\theta_1$ and the neutrino 
masses $m_i$ are unrestricted by \eq{23sym}, as they can be adjusted by the 
choice of the four parameters $a,b,f,e$. Hence it is possible to 
accommodate the SMA, the LMA, the LOW, or any other solution of solar
neutrino oscillations.

If a fundamental symmetry like \eq{23sym} is present, 
most likely it occurs at a high-energy scale 
$E_N$ set by the right-handed neutrinos, and not at present
experimental energies.
So, strictly speaking, a renormalization-group correction must
be applied to \eq{u0} before comparing it with experiments.
However, if the correction is small, then the 2-3 symmetry is
still approximately valid at present energies. This is indeed the case.
The corrected result agrees with experimental observations,
{\it for certain scenarios but not for others}. Consequently, if we 
believe the 2-3 symmetry to be a valid high-energy symmetry, then it
can also be used to rule out possible scenarios not directly refuted by
current experiments.

The neutrino mass matrix at energy scale $E_N$ that possesses the 2-3
symmetry will continue to be denoted by $m'$. The renormalization-group
corrected neutrino mass matrix at experimental energies will be denoted
by $M'$. It is known that they are related by \cite{RG}
\be (M')_{\alpha\beta}=\sqrt{I_\alpha I_\beta}(m')_{\alpha\beta},
\quad (\alpha,\beta=e,\mu,\tau).
\labels{mlow}\ee
The quantity $I_\alpha$ is determined
from the charged-lepton Yukawa coupling $h_\alpha$ by the relation
\be
I_\alpha=\exp[-\int_{t}^{t_N} dt'\ h^2_\alpha(t')/8\pi^2],\labels{i}\ee
where $t$ is the logarithm of the present energy and $t_N$
is the logarithm of $E_N$.
With \eq{mlow},
the symmetry requirement \eq{23sym} is  weakened to read
\be
\({M'_{e\tau}\over M'_{e\mu}}\)^2={M'_{\tau\tau}\over M'_{\mu\mu}}=
{I_\tau\over I_\mu}.
\labels{ratio}\ee 
Although there are still two relations, the second one is now
a restriction on the ratio imposed by
the renormalization-group.
As a result, the MNS matrix is no longer given by \eq{u0}, and
a deviation from maximal atmospheric mixing and $u_{e3}=0$ will occur.
Nevertheless,
since $I_\alpha$ are generally quite close to 1, the
deviations, though present, are small and are consistent with a global
fit of the data \cite{GGMPV}. With a 2-3 symmetry, these small
deviations are attributed to renormalization-group effects.

Following Ref.~\cite{EL}, we assume the renormalization-group dynamics
to be determined by MSSM with the following parameters:
the right-handed neutrino scale  $E_N=10^{13}$ GeV,
the GUT unification scale  $M_{GUT}=1.1\x 10^{16}$ GeV, with
a coupling $\alpha_{GUT}^{-1}=25.64$, 
and the SUSY scale  $M_{SUSY}=1$ TeV.
The Yukawa couplings are taken to be
$h_{top}=3.0$, and $h_{bottom},h_\tau, h_\mu, h_e$ 
are determined by the observed fermion masses. With this choice,
 $I_e$ is always very close
to 1 and will be taken as such. 
$I_\tau$ varies from 0.826 at $h_\tau=3.0$ (corresponding to $\tan\beta=58.2)$
monotonically to 0.99997 at $h_\tau=0.013\ (\tan\beta=1)$, while
$I_\mu$ varies monotonically from 0.9955 to 1.00000
\cite{EL}. In this range, the ratio $I_\tau/I_\mu$ needed in
\eq{ratio} varies from 0.8297
to 0.99997. Ratios beyond this range will be rejected.

The following procedure is employed to test the hypothesis \eq{ratio},
or equivalently, \eq{23sym}.
Using upper-case letters to denote \eq{mpum} at experimental energies,
we compute $M'=U^TMU$ from each of the solutions allowed by the
global fit in Ref.~\cite{GGMPV}. The details of this will be discussed
in the next paragraph.
Then we test whether \eq{ratio}
is obeyed in two steps. First, we vary $\tan^2\theta_3$ between 0
and the approximate upper bound 0.026 set by the CHOOZ reactor experiment
\cite{CHOOZ} 
to achieve the first equality in \eq{ratio}. If that is not possible
the solution is rejected, and a cross ($\x$) is put in Table I.
If the solution passes this test, then the value of $\tan^2\theta_3$
is recorded in Table I, and we proceed to compute $I_\tau/I_\mu$ from
the second equality in \eq{ratio}. If the value does not lie between
and 0.8297 and 0.99997, we reject the solution and put a cross in Table
I. Otherwise the value is recorded in Table I. 

The following parameters from atmospheric neutrino experiments
will be fixed throughout \cite{GGMPV}: $\tan^2\theta_2
=1.6$, and $\Delta m_{23}^2=3\x 10^{-3}$ eV$^2$. 
For solar neutrino observations, we investigate all three viable
solutions \cite{GGMPV}, SMA, LMA, and LOW,
with $(\tan^2\theta_1, \Delta m_{12}^2)$ respectively equal to
to $(8\x 10^{-4},\
5\x10^{-6}), (0.4, 3\x10^{-5})$, and $(0.8, 10^{-7})$, and
 $\Delta m_{12}^2$ expressed in 
eV$^2$. In each of these three cases, 
the only remaining parameter in $U$ is $tan^2\theta_3$.
As for $M$, we shall
consider two scenarios in each case: the normal hierarchy
(nh), where $0\le m_0\equiv m_1<m_2<m_3$, and the inverted hierarchy
(ih),  where $m_1>m_2>m_3\equiv m_0\ge 0$. 
In each of these two scenarios the observed $\Delta m_{12}^2$ and 
$\Delta m_{23}^2$ leaves only one free parameter $m_0$ in the diagonal mass
matrix $M$. 
So in toto, we have six
cases to consider, (SMAnh), (SMAih), (LMAnh), (LMAih), (LOWnh), (LOWih),
each of them having two adjustable parameters, $m_0$
and $\tan^2\theta_3$. As discussed in the last paragraph, these two
parameters are adjusted so that \eq{ratio} holds. Since $I_\tau/I_\mu$
can vary over a range, depending on the value of $\tan\beta$, effectively
there is one free parameter left, which is taken to be the minimal
neutrino mass $m_0$ in Table I. An upper cutoff of 2 eV is also imposed
on the Table, roughly corresponds to the bound observed in the Mainz experiment
\cite{MAINZ}. One may also want to read the Table  with a more stringent
upper bound of 0.26 eV in mind, correponding to the experimental limit
of neutrinoless double beta decay from the Heidelberg-Moscow experiment
\cite{KK}.

The result is summarized in Table I, for each of
the six cases, and a range of $m_0$ between 0 and 2 eV. 
We see there that the inverse mass hierarchy
(ih) is ruled out by the 2-3 symmetry \eq{23sym}, but
with normal mass hierarchy (nh), all three solar neutrino solutions
(SMA, LMA, LOW) are allowed. The most striking result of the
2-3 symmetry is that there is a {\it lower bound}  
around 0.025 eV for the neutrino masses. This lowerbound increases when
$\tan\beta$ decreases from 58.2, which corresponds to $h_\tau=h_{top}=3.0$.
We also see that 
$\tan^2\theta_3$ are always very small ($\sim 10^{-6}$ or $\sim 10^{-7}$) 
for the SMA and LOW solutions; only the LMA solution gives a relatively 
large  $\tan^2\theta_3$, of the order $10^{-2}$ to $10^{-3}$.

The 2-3 symmetry has been assumed to hold for the left-handed neutrino
mass matrix \eq{mp}, in the basis where the charged leptons are
mass diagonal. By itself the symmetry makes no statement about how right-handed
leptons transform, nor even how the left-handed charged leptons behave
under a 2-3 transformation. If one assumes that the left-handed
charged leptons must transform the same way as the left-handed neutrinos,
then in order to keep the charged leptons mass diagonal, 
these  mass terms must break the 2-3 symmetry and the
right-handed charged leptons must not transform like the left-handed ones.
The dynamical mechanism which could bring this about is 
presently not understood.

To summarize, neutrino oscillation data with a normal mass hierarchy
(nh) are consistent with a 
2-3 symmetry \eq{23sym}, at the right-handed neutrino
scale of $E_N\sim 10^{13}$ GeV and a renormalization-group extrapolation
using MSSM. This symmetry predicts a minimum value for the electron
neutrino mass around 0.025 eV. The actual bound
depends on the value of $\tan\beta$; it increases with decreasing
$\tan\beta$. The values of $\tan^2\theta_3$ are always tiny for
the SMA and the LOW solutions, and they become somewhat substantial
($\sim 10^{-2}$ to $10^{-3}$) only in the LMA solution.

This research is supported by the Natural Science and Engineering Research
Council of Canada and the Qu\'ebec Department of Education.

\newpage
\begin{table}
\begin{tabular}{|c|l|l|l|} \hline 
Type&$m_0$\ (eV)&$\tan^2\theta_3$&$I_\tau/I_\mu$\\ \hline
SMAnh&0.00&0.016&$\x$\\ 
&     0.02&$2.3\x10^{-6}$&$\x$\\
&     0.02565&$1.3\x10^{-6}$&0.8298\\
&     0.03&$9.0\x10^{-7}$&0.8505\\
&     0.04&$5.3\x10^{-7}$&0.8878\\
&     0.05&$3.8\x10^{-7}$&0.9142\\
&     0.10&$2.1\x10^{-7}$&0.9702\\
&     0.50&$1.6\x10^{-7}$&0.9986\\ 
&     2.00&$1.6\x10^{-7}$&0.9999\\ \hline
SMAih&0.00&$2.1\x10^{-13}$&$\x$\\
&     2.00&$1.6\x10^{-7}$&$\x$\\ \hline
\end{tabular}

\bigskip\bigskip
\begin{tabular}{|c|l|l|l|} \hline
Type&$m_0$\ (eV)&$\tan^2\theta_3$&$I_\tau/I_\mu$\\ \hline
LMAnh&0.00&$\x$&\\
&     0.01&$\x$&\\
&     0.02536&0.0117&0.8298\\
&     0.035&0.0061&0.8715\\
&     0.05&0.0035&0.9143\\
&     0.10&0.0019&0.9702\\
&     0.50&0.0015&0.9986\\
&     1.00&0.0015&0.9997\\
&     2.00&0.0015&0.9999\\ \hline
LMAih&0.00&$8.8\x10^{-14}$&$\x$\\
&     2.00&0.0015&$\x$\\ \hline
\end{tabular}

\bigskip\bigskip
\begin{tabular}{|c|l|l|l|} \hline
Type&$m_0$\ (eV)&$\tan^2\theta_3$&$I_\tau/I_\mu$\\ \hline
LOWnh&0.00&$\x$&\\ \hline
&     0.001&0.0108&$\x$\\
&	0.02575&$1.6\x10^{-7}$&0.8298\\
&	0.05&$4.8\x10^{-8}$&0.9140\\
&     0.10&$2.6\x10^{-8}$&0.9702\\
&     1.00&$2.0\x10^{-8}$&0.9997\\
&     2.00&$2.0\x10^{-8}$&0.9999\\ \hline
LOWih&0.00&0&$\x$\\
&     2.00&$2\x10^{-8}$&$\x$\\ \hline         
\end{tabular}

\bigskip
\caption{Allowed values of the minimal neutrino mass $m_0$, and the 
solutions of $\tan^2\theta_3$ and $I_\tau/I_\mu$. In the MSSM dynamics
considered in this paper,
the last column
decreases from 0.9997 at $\tan\beta=1$, to 0.8297 at $\tan\beta=58.2$,
where $m_{top}=m_{tau}$ at $M_{GUT}$. A cross ($\x$) indicates the absence of
a solution at that value of $m_0$. The first subtable
deals with the solar solution SMA, with both normal mass hierarchy (nh)
and inverse mass hierarchy (ih). The other two subtables deal with
similar cases for the solar solutions LMA and LOW.}

\end{table}

\end{document}